\documentclass[twocolumn,journal]{IEEEtran}
\PassOptionsToPackage{ruled}{algorithm2e}
\usepackage[T1]{fontenc}
\usepackage[latin9]{inputenc}
\usepackage{color}
\usepackage{refstyle}
\usepackage{algorithm2e}
\usepackage{amsmath}
\usepackage{amsthm}
\usepackage{amssymb}
\usepackage{graphicx}
\usepackage{wasysym}
\usepackage[unicode=true,
 bookmarks=true,bookmarksnumbered=true,bookmarksopen=true,bookmarksopenlevel=1,
 breaklinks=false,pdfborder={0 0 0},pdfborderstyle={},backref=false,colorlinks=false]
 {hyperref}
\hypersetup{pdftitle={Your Title},
 pdfauthor={Your Name},
 pdfpagelayout=OneColumn, pdfnewwindow=true, pdfstartview=XYZ, plainpages=false}

\makeatletter

\AtBeginDocument{\providecommand\algref[1]{\ref{alg:#1}}}
\AtBeginDocument{\providecommand\lineref[1]{\ref{line:#1}}}
\AtBeginDocument{\providecommand\thmref[1]{\ref{thm:#1}}}
\RS@ifundefined{subsecref}
  {\newref{subsec}{name = \RSsectxt}}
  {}
\RS@ifundefined{thmref}
  {\def\RSthmtxt{theorem~}\newref{thm}{name = \RSthmtxt}}
  {}
\RS@ifundefined{lemref}
  {\def\RSlemtxt{lemma~}\newref{lem}{name = \RSlemtxt}}
  {}

\theoremstyle{plain}
\newtheorem{thm}{\protect\theoremname}

\usepackage[caption=false,font=footnotesize]{subfig}

\DeclareMathOperator{\minimize}{minimize}
\DeclareMathOperator{\st}{subject~to}

\DeclareMathOperator{\tr}{Tr}

\DeclareMathOperator{\vect}{vec}

\newref{alg}{refcmd = \textbf{Algorithm~\ref{#1}}}
\newref{line}{refcmd = \textbf{\ref{#1}}}
\newref{thm}{refcmd = \textbf{Theorem \ref{#1}}}

\usepackage{cite}
\usepackage{acronym}
\AtBeginDocument{\acrodef{AO}{alternating optimization}
\acrodef{APGM}{alternating projected gradient method}
\acrodef{APM}{accelerated proximal gradient method}
\acrodef{ASP}{antenna separation product}
\acrodef{AWGN}{additive white Gaussian noise}
\acrodef{BC}{broadcast channel}
\acrodef{BCM}{block coordinate maximization}
\acrodef{BEP}{bit error probability}
\acrodef{BER}{bit error rate}
\acrodef{BF-MIMO}[BF\mbox{-}MIMO]{beamforming MIMO}
\acrodef{BF}{beamforming}
\acrodef{BS}{base station}
\acrodef{bpcu}{bits per channel use}
\acrodef{CP}{cyclic prefix}
\acrodef{CR}{cutoff rate}
\acrodef{CSI}{channel state information}
\acrodef{CSIR}{channel state information at RX}
\acrodef{SSK}{space shift keying}
\acrodef{CSIT}{channel state information at TX}
\acrodef{DCMC}{discrete\mbox{-}input continuous\mbox{-}output memoryless channel}
\acrodef{DFT}{discrete Fourier transform}
\acrodef{DL-TR-GSM}{dual-layered transmit-receive \acl{GSM}}
\acrodef{DLT}{dual-layered transmission}
\acrodef{DOA}{direction of arrival}
\acrodef{DPC}{dirty paper coding}
\acrodef{EGC}{equal gain combining}
\acrodef{EM}{electromagnetic}
\acrodef{EVD}{eigenvalue decomposition}
\acrodef{FPGA}{field programmable gate array}
\acrodef{FSPL}{free space path loss}
\acrodef{FFT}{fast Fourier transform}
\acrodef{FDE}{frequency domain equalization}
\acrodef{GRSM}{generalized \acl{RSM}}
\acrodef{GSM}{generalized \acl{SM}}
\acrodef{HMIMO}{holographic MIMO}
\acrodef{IFFT}{invserse fast Fourier transform}
\acrodef{ICI}{inter-channel interference}
\acrodef{iid}[i.i.d.]{independent and identically distributed}
\acrodef{IQ}{in\mbox{-}phase and quadrature}
\acrodef{ISI}{intersymbol interference}
\acrodef{ISI-free}[ISI\mbox{-}free]{intersymbol interference free}
\acrodef{LIS}{large intelligent surface}
\acrodef{LOS}{line\mbox{-}of\mbox{-}sight}
\acrodef{KKT}{Karush\mbox{-}Kuhn\mbox{-}Tucker} 
\acrodef{MAC}{multiple-access channel}
\acrodef{mmWave}{millimeter-wave}
\acrodef{MI}{mutual information}
\acrodef{MIMO}{multiple\mbox{-}input multiple\mbox{-}output}
\acrodef{mMIMO}{massive MIMO}
\acrodef{MISO}{multiple\mbox{-}input single\mbox{-}output}
\acrodef{ML}{maximum likelihood}
\acrodef{MRC}{maximal ratio combining}
\acrodef{MMSE}{minimum mean square error}
\acrodef{MU-TR-GSM}{multiuser transmit-receive  \acl{GSM} }
\acrodef{NCSIT}{no channel state information at TX}
\acrodef{NLOS}{non\mbox{-}\acs{LOS}} 
\acrodef{NOMA}{non-orthogonal multiple access}
\acrodef{OFDM}{orthogonal frequency division multiplexing}
\acrodef{OFDMA}{orthogonal frequency division multiple access}
\acrodef{PA}{power amplifier}
\acrodef{PAE}{power added efficiency}
\acrodef{PAPR}{peak\mbox{-}to\mbox{-}average power ratio}
\acrodef{PDF}{probability density function}
\acrodef{PEP}{pairwise error probability}
\acrodefplural{PEP}{pairwise error probabilities}
\acrodef{PGM}{projected gradient method}
\acrodef{PMP}{probability mass function}
\acrodef{PSM}{precoding-aided spatial modulation}
\acrodef{QSM}{quadrature spatial modulation}
\acrodef{RC}{reorganization computation}
\acrodef{RHS}{right-hand side}
\acrodef{RIS}{reconfigurable intelligent surface}
\acrodef{RSM}{receive spatial modulation}
\acrodef{RX}{receiver}
\acrodef{SEP}{symbol error probability}
\acrodef{SER}{symbol error rate}
\acrodef{SIC}{successive interference cancellation}
\acrodef{SIM}{stacked intelligent metasurface}
\acrodef{SINR}{signal-to-interference-plus-noise ratio}
\acrodef{SISO}{single-input single-output}
\acrodef{SM}{spatial modulation}
\acrodef{SMX-MIMO}[SMX\mbox{-}MIMO]{spatial multiplexing MIMO}
\acrodef{SMX}{spatial multiplexing}
\acrodef{SNR}{signal-to-noise ratio}
\acrodef{SC}{single carrier}
\acrodef{SVD}{singular value decomposition}
\acrodef{SPST}{single pole single-throw}
\acrodef{SU}{secondary user}
\acrodef{TDE}{time domain equalization}
\acrodef{TX}{transmitter}
\acrodef{ULA}{uniform linear array}
\acrodef{URA}{uniform rectangular array}
\acrodef{VGA}{variable gain amplifier}
\acrodef{wrt}[w.r.t.]{with respect to}
\acrodef{ZF}{zero-forcing}
\acrodef{ZMCG}{zero-mean complex Gaussian}

}

\makeatother

\providecommand{\theoremname}{Theorem}

\begin{document}
\title{Mutual Information Optimization for SIM-Based Holographic MIMO Systems}
\author{Nemanja Stefan Perovi\'c,~\IEEEmembership{Member,~IEEE} and Le-Nam
Tran, \IEEEmembership{Senior Member, IEEE}\thanks{This publication has emanated from research supported in part by research
grants from Science Foundation (SFI) under Grant Numbers 22/US/3847
and 17/CDA/4786.}\thanks{N.~S.~Perovi\'c was with Universit\'e Paris-Saclay, CNRS, CentraleSup\'elec,
Laboratoire des Signaux et Syst\`emes, 3 Rue Joliot-Curie, 91192
Gif-sur-Yvette, France. Email: n.s.perovic@gmail.com.}\thanks{L.-N. Tran is with the School of Electrical and Electronic Engineering,
University College Dublin, Belfield, Dublin 4, D04~V1W8, Ireland.
Email: nam.tran@ucd.ie.}}
\maketitle
\begin{abstract}
In the context of emerging \ac{SIM}-based \ac{HMIMO} systems, a
fundamental problem is to study the \ac{MI} between transmitted and
received signals to establish their capacity. However, direct optimization
or analytical evaluation of the MI, particularly for discrete signaling,
is often intractable. To address this challenge, we adopt the channel
\ac{CR} as an alternative optimization metric for the \ac{MI} maximization.
In this regard, we propose an \ac{APGM}, which optimizes the \ac{CR}
of a \ac{SIM}-based \ac{HMIMO} system by adjusting signal precoding
and the phase shifts across the transmit and receive \acp{SIM} on
a layer-by-layer basis. \textcolor{black}{Simulation results indicate
that the proposed algorithm significantly enhances the CR, achieving
substantial gains, compared to the case with random SIM phase shifts,
that are proportional to those observed for the corresponding MI.
}This justifies the effectiveness of using the channel \ac{CR} for
the \ac{MI} optimization. Moreover, we demonstrate that the integration
of digital precoding, even on a modest scale, has a significant impact
on the ultimate performance of SIM-aided systems. \acresetall{}
\end{abstract}

\begin{IEEEkeywords}
Channel \ac{CR}, \ac{MI}, \ac{SIM}, \ac{HMIMO}, optimization.\acresetall{}\vspace{-0.3cm}
\end{IEEEkeywords}

\section{Introduction}

\bstctlcite{BSTcontrol}Intelligent metasurfaces are expected to significantly
impact the evolution of future wireless communications. These surfaces,
composed of a large number of controllable metamaterial elements,
can dynamically modify the \ac{EM} wave propagation \cite{di2020smart}.
This ability enables them to improve energy \textcolor{black}{efficiency
\cite{wang2023energy}, network capacity, reliability \cite{zhu2023robust}
and connectivity \cite{lin2022refracting},} while also supporting
other functionalities, such as localization and sensing. However,
their advantages hinge on acquiring accurate \ac{CSI}, a challenging
task for conventional structures, such as \acp{RIS}. Furthermore,
the path loss in the \ac{RIS}-assisted links can substantially diminish
these benefits, indicating that placing intelligent metasurfaces close
to transmitters and receivers maximizes their potential.

The above discussions naturally motivate the integration of intelligent
metasurface structures with wireless communication transceivers in
\ac{mMIMO} and \ac{HMIMO} \textcolor{black}{systems \cite{zhu2024beamforming}.}
The use of intelligent metasurfaces in these systems offers a tradeoff
between minimizing the number of RF chains and increasing control
over \ac{EM} fields \cite{di2024electromagnetic}. Moreover, they
can provide beamforming capabilities comparable to those of conventional
phased array antennas, but at lower cost and power consumption. However,
limitations due to their typically single-layer structure and the
finite number of tunable states of their elements in practice can
result in beam misalignment, potentially compromising the expected~performance.

To address the above limitations, multi-layer metasurface structures
have emerged as a promising solution for flexibly forming different
radiation patterns compared to their single-layer \textcolor{black}{counterparts
\cite{an2024exploiting}.} Such structures, called \acp{SIM}, were
recently introduced \cite{an2023stacked}, and their origin was inspired
from the architecture of a deep neural \textcolor{black}{network \cite{liu2022programmable}.}
Indeed, \acp{SIM} mirror the multi-layer structure of neural networks
that are capable of modeling various functions with improved signal
processing capabilities. Similarly, when metasurface layers are properly
placed, \ac{SIM} can implement signal processing directly in the
\ac{EM} wave domain. This approach can potentially reduce the reliance~on
digital beamforming and high-precision digital-to-analog converters.

In \cite{an2024two}, the signal processing capabilities of SIMs were
exploited for the implementation of 2D \ac{DFT} for \ac{DOA} estimation.
Moreover, a hybrid channel estimator, in which the received training
symbols were processed first in the wave domain and later in the digital
domain, was proposed in \cite{nadeem2023hybrid}. In \cite{an2023stackedmulti},
the authors studied the achievable sum-rate maximization for a downlink
channel between a SIM-assisted base station and multiple single-antenna
users. Integrating SIMs with transmitters and receivers into the so-called
SIM-based HMIMO system, which performs signal precoding and combining
in the wave domain, was proposed in \cite{an2023stackedholo}. Moreover,
the achievable rate optimization for the SIM-based HMIMO system was
studied in \cite{papazafeiropoulos2024achievable}.

Despite quite a few papers about SIM-based systems, none of them has
considered the achievable rate of such systems using discrete signaling,
i.e., the \ac{MI}. This gap has motivated us to investigate the optimization
of the MI in the aforementioned systems. As shall be seen shortly,
the direct optimization of the MI is intractable, and thus we instead
propose the use of the channel \ac{CR} as an alternative metric to
facilitate the MI optimization \cite{perovic2018optimization,perovic2021optimization},
which is justified as follows\textcolor{black}{. First, the MI is
related to the CR, $R_{0}$, by $\text{MI}\geq N_{r}(1-\log_{2}e)+R_{0}$
\cite[Eqn. (36)]{perovic2018optimization}, meaning that maximizing
the CR accordingly translates into maximizing the MI. Second, $R_{0}$
and the codeword error probability, $P_{e}$, are related by $P_{e}\le2^{-n(R_{0}-R)}$
\cite[Eq. (6.8\textendash 19)]{john2008digital}, where $n$ is the
codeword length and $R$ is the information rate. Essentially, letting
$n\rightarrow\infty$, $P_{e}$ can be made arbitrarily small provided
that $R<R_{0}$. Since the channel capacity represents the theoretical
maximum information rate for reliable communications, the CR is effectively
employed as a practical lower bound on the channel capacity. The inter-connection
among the CR, the MI, and the channel capacity indeed suggests that
the CR can be used to optimize the MI in an indirect, yet efficient,
way.}

The main contributions of this paper are listed as follows:
\begin{itemize}
\item We show that using the channel \ac{CR} as an alternative optimization
metric enables efficient maximization of the MI in SIM-based HMIMO
systems. Specifically, the CR is expressed as a closed-form expression,
which facilitates the derivation of an efficient optimization algorithm.
\item To maximize the CR, we formulate a joint optimization problem of transmit
precoding and the phase shifts for both the transmit and the receive
SIMs. An \ac{APGM} is proposed to solve this problem, which particularly
optimizes the phase shifts for the transmit and the receive SIMs on
a layer-by-layer basis\textcolor{black}{, using an independent step
size for each layer, which is different from \cite{an2023stackedholo}.}
Also, we provide the computational complexity of the proposed algorithm.
\item We demonstrate through simulation results that the proposed algorithm
can significantly increase the CR and the MI in a proportional manner.
Also, the increase of the modulation order without changing the transmit
signal power has a negligible influence on the MI. Finally, incorporating
even a small scale digital precoding into a SIM-based HMIMO system
leads to substantial improvements on the achievable MI.
\end{itemize}
\textcolor{black}{\emph{Notation}}\textcolor{black}{: Bold lower and
upper case letters represent vectors and matrices, respectively. $\tr(\cdot)$
denotes the trace of a matrix and $\text{diag}(\cdot)$ transforms
a vector into a diagonal matrix. $\log_{2}(\cdot)$ is the binary
logarithm, $\exp(\cdot)$ denotes the exponent, $\mathbb{E}\{\cdot\}$
denotes the expectation operation and $(\cdot)^{*}$ is the complex
conjugate. $(\cdot)^{T}$ and $(\cdot)^{H}$ represent transpose and
Hermitian transpose, receptively. $\nabla_{x}f(\cdot)$ is the complex
gradient of $f(\cdot)$ \ac{wrt} $x^{*}$.}\textcolor{blue}{{} }\vspace{-0.2cm}

\section{System Model}

We consider a SIM-based HMIMO communication system with $N_{t}$ transmit
antennas and $N_{r}$ receive antennas, where both the transmitter
and the receiver are equipped with SIMs. The transmit SIM consists
of $L$ metasurface layers with $N$ meta-atoms in each layer, while
the receive SIM consists of $K$ metasurface layers with $E$ meta-atoms
in each layer. \textcolor{black}{We assume that the signal propagation
through every SIM layer is ideal, i.e., without any power loss.} The
SIM meta-atoms are connected to smart controllers, which can independently
adjust the phase shift of each meta-atom. In this way, MIMO transceivers
with closely spaced SIMs~can implement signal beamforming directly
in the EM wave domain.

For the $l$-th layer of the transmit SIM, the phase shift matrix
is denoted as $\boldsymbol{\mathbf{\Phi}}^{l}=\text{diag}(\boldsymbol{\phi}^{l})=\text{diag}([\phi_{1}^{l}\:\phi_{2}^{l}\:\cdots\:\phi_{N}^{l}]^{T})$.
Also, $\phi_{n}^{l}=\exp(j\theta_{n}^{l})$, where $\theta_{n}^{l}$
is the phase shift of the $n$-th meta-atom in the $l$-th layer.
Signal propagation between the $(l-1)$-th and $l$-th layer of the
transmit SIM is modeled by the matrix $\mathbf{W}^{l}\in\mathbb{C}^{N\times N}$,
where $[\mathbf{W}^{l}]_{m,n}$ denotes the signal propagation between
the $n$-th meta-atom of the $(l-1)$-th layer and the $m$-th meta-atom
of the $l$-th layer. According to Rayleigh-Sommerfeld diffraction
theory, $[\mathbf{W}^{l}]_{m,n}$ is equal \cite[Eq. (1)]{lin2018all}
\begin{equation}
[\mathbf{W}^{l}]_{m,n}=\frac{A\cos\chi_{m,n}}{d_{m,n}}\Bigl(\frac{1}{2\pi d_{m,n}}-\frac{j}{\lambda}\Bigr)e^{j\frac{2\pi d_{m,n}}{\lambda}}\label{eq:RS_equ}
\end{equation}
for $l=2,3,\dots,L$, where $A$ is the area of each \textcolor{black}{meta-atom,
$\lambda$ is the wavelength, }$d_{m,n}$ is the propagation distance
between the meta-atoms in the $(l-1)$-th and $l$-th layer of the
transmit SIM, $\chi_{m,n}$ is the angle between the propagation direction
and normal direction of the $(l-1)$-th layer\footnote{In this paper, it is assumed that all layers of the transmit SIM have
the same arrangement and thus the \ac{RHS} of (\ref{eq:RS_equ})
does not depend on $l$. The same applies to the receive SIM.}. Also, signal propagation between the transmit antenna array and
the first layer of the transmit SIM is modeled by $\mathbf{W}^{1}\in\mathbb{C}^{N\times N_{t}}$,
which is defined as (\ref{eq:RS_equ}). The wave domain beamforming
matrix of the transmit SIM is given as $\mathbf{B}=\boldsymbol{\mathbf{\Phi}}^{L}\mathbf{W}^{L}\cdots\boldsymbol{\mathbf{\Phi}}^{2}\mathbf{W}^{2}\boldsymbol{\mathbf{\Phi}}^{1}\mathbf{W}^{1}\in\mathbb{C}^{N\times N_{t}}.$

At the receive SIM, the phase shift matrix for the $k$-th layer is
given by $\boldsymbol{\mathbf{\Psi}}^{k}=\text{diag}(\text{\ensuremath{\boldsymbol{\psi}}}^{k})=\text{diag}([\psi_{1}^{k}\:\psi_{2}^{k}\:\cdots\:\psi_{E}^{k}]^{T})$,
where $\psi_{e}^{k}=\exp(j\upsilon_{e}^{k})$ and $\upsilon_{e}^{k}$
is the phase shift of the $e$-th meta-atom in the $k$-th layer.
Signal propagation between the $k$-th and $(k-1)$-th layer of the
receive SIM is modeled by the matrix $\mathbf{U}^{k}\in\mathbb{C}^{E\times E}$,
which is defined similarly as in (\ref{eq:RS_equ}). Similarly, signal
propagation between the first layer of the receive SIM and the receive
antenna array is given by $\mathbf{U}^{1}\in\mathbb{C}^{N_{r}\times E}$.
The wave domain beamforming matrix for the receive SIM can be expressed
as $\text{\ensuremath{\mathbf{Z}}}=\mathbf{U}^{1}\boldsymbol{\Psi}^{1}\mathbf{U}^{2}\boldsymbol{\Psi}^{2}\cdots\mathbf{U}^{K}\boldsymbol{\Psi}^{K}\in\mathbb{C}^{N_{r}\times E}.$

Let $\mathbf{G}\in\mathbb{C}^{N\times E}$ be the channel between
the transmit and the receive SIM. Then the end-to-end channel matrix
for the considered communication system is $\mathbf{H}=\ensuremath{\mathbf{Z}}\ensuremath{\mathbf{G}}\ensuremath{\mathbf{B}}\in\mathbb{C}^{N_{r}\times N_{t}}$
and the signal vector at the receive antennas is given by 
\begin{equation}
\mathbf{y}=\mathbf{H}\mathbf{P}\mathbf{x}_{i}+\mathbf{n}
\end{equation}
where $\mathbf{P}\in\mathbb{C}^{N_{t}\times N_{s}}$ is the precoding
matrix that satisfies the following the average transmit power constraint:
\begin{equation}
\tr(\mathbf{P}\mathbf{P}^{H})=N_{s}\label{eq:P_constr}
\end{equation}
where $N_{s}\le\min(N_{t},N_{r})$ is the number of the transmitted
modulation symbols. The transmit vector $\mathbf{x}_{i}\in\mathbb{C}^{N_{s}\times1}$
consists of elements chosen from a discrete symbol alphabet of size
$M$, and thus, the number of different transmit vectors is \mbox{$N_{vec}=M^{N_{s}}$.}
In addition, it is assumed that the average symbol energy of the discrete
symbol alphabet is one. Finally, $\mathbf{n}\in\mathbb{C}^{N_{r}\times1}$
is the noise vector with \ac{iid} elements that are distributed according
to $\mathcal{CN}(0,\sigma^{2})$, where $\sigma^{2}$ is the noise
variance. \vspace{-0.2cm}

\section{Problem Formulation}

We consider a \ac{DCMC}. For equally probable transmitted symbol
vectors, the \ac{MI} is found as \cite{perovic2018optimization,perovic2021optimization}
\begin{equation}
\text{MI}=\log_{2}(N_{vec})-\frac{1}{N_{vec}}\sum\nolimits _{i=1}^{N_{vec}}\mathbb{E}_{\mathbf{n}}\Bigl\{\log_{2}\sum\nolimits _{j=1}^{N_{vec}}e^{\kappa_{i,j}}\Bigr\}\label{eq:MI}
\end{equation}
where $\kappa_{i,j}=(-||\mathbf{H}\ensuremath{\mathbf{P}}(\mathbf{x}_{i}-\mathbf{x}_{j})+\mathbf{n}||^{2}+||\mathbf{n}||^{2})/\sigma^{2}$.
As the direct optimization of the MI in (\ref{eq:MI}) involves discrete
variables, which is intractable, we instead consider the optimization
of a closed related metric, known as the \ac{CR}, which is given
by
\begin{equation}
R_{0}=-\log_{2}\Bigl[\frac{1}{N_{vec}^{2}}\sum\nolimits _{i,j=1}^{N_{vec}}e^{-\frac{F_{i,j}}{4\sigma^{2}}}\Bigr]\label{eq:CR}
\end{equation}
where $F_{i,j}=||\mathbf{H}\ensuremath{\mathbf{P}}(\mathbf{x}_{i}-\mathbf{x}_{j})||^{2}=||\mathbf{H}\ensuremath{\mathbf{P}}\APLup\mathbf{x}_{i,j}||^{2}$.
Since the CR decreases with the argument of the logarithm function
in (\ref{eq:CR}), maximizing $R_{0}$ is equivalent to the following
optimization problem \cite{perovic2021optimization}
\begin{subequations}
\label{eq:Opt_problem}
\begin{align}
\text{\ensuremath{\underset{\mathbf{P},\boldsymbol{\phi},\boldsymbol{\psi}}{\minimize}}	} & f(\mathbf{P},\boldsymbol{\phi},\boldsymbol{\psi})=\sum\nolimits _{i,j=1}^{N_{vec}}e^{-\frac{F_{i,j}}{4\sigma^{2}}}\label{eq:obj_fun}\\
\st\text{	} & \bigl|\phi_{n}^{l}\bigr|=1,\;\forall l,n\label{eq:phi-modulus}\\
 & \bigl|\psi_{e}^{k}\bigr|=1,\;\forall k,e\label{eq:psi-modulus}\\
 & \tr(\mathbf{P}\mathbf{P}^{H})=N_{s},
\end{align}
\end{subequations}
where $\boldsymbol{\phi}=[(\boldsymbol{\phi}^{1})^{T},(\boldsymbol{\phi}^{2})^{T},\ldots,(\boldsymbol{\phi}^{L})^{T}]^{T}\in\mathbb{C}^{(NL)\times1}$,
$\boldsymbol{\boldsymbol{\psi}}=[(\boldsymbol{\psi}^{1})^{T},(\boldsymbol{\psi}^{2})^{T},\ldots,(\boldsymbol{\psi}^{K})^{T}]^{T}\in\mathbb{C}^{(MK)\times1}$,
and the equalities in (\ref{eq:phi-modulus}) and (\ref{eq:psi-modulus})
are treated element-wise.\vspace{-0.2cm}

\section{Proposed Optimization Method}

\begin{algorithm}[t]
{\small

\SetAlgoNoLine
\DontPrintSemicolon
\LinesNumbered 

\KwIn{$\mathbf{P}_{0}$, $\boldsymbol{\phi}_{0}$, $\boldsymbol{\psi}_{0}$,
$\nu_{0}>0$, $\mu_{0}^{1:L}>0$, $\tau_{0}^{1:K}>0$, $n\leftarrow0$,
$0<\rho<1$, $\delta>0$.}

\Repeat{convergence}{

\Repeat(\tcc*[f]{line search for ${\mathbf{P}}$}){\label{line:PLS:end} $f(\mathbf{P}_{n+1},\boldsymbol{\phi}_{n},\boldsymbol{\psi}_{n})\le f(\mathbf{P}_{n},\boldsymbol{\phi}_{n},\boldsymbol{\psi}_{n})-\delta\bigl\Vert\mathbf{P}_{n+1}-\mathbf{P}_{n}\bigr\Vert^{2}$}{\label{line:PLS:start}

$\mathbf{P}_{n+1}=\mathcal{P}_{P}(\mathbf{P}_{n}-\nu_{n}\nabla_{\mathbf{P}}f(\mathbf{P}_{n},\boldsymbol{\phi}_{n},\boldsymbol{\psi}_{n}))$\;\label{alg:APGM:pgS}

\If{$f(\mathbf{P}_{n+1},\boldsymbol{\phi}_{n},\boldsymbol{\psi}_{n})>f(\mathbf{P}_{n},\boldsymbol{\phi}_{n},\boldsymbol{\psi}_{n})-\delta\bigl\Vert\mathbf{P}_{n+1}-\mathbf{P}_{n}\bigr\Vert^{2}$
}{$\nu_{n}\leftarrow\rho\nu_{n}$\;}

}

\For{$l=1$ $\boldsymbol{\mathrm{to}}$ $L$ }{

\Repeat(\tcc*[f]{line search for $\boldsymbol{\phi}^{l}$}){\label{line:phiLS:end}$f(\mathbf{P}_{n+1},\bar{\boldsymbol{\phi}}_{n+1}^{l},\boldsymbol{\psi}_{n})\le f(\mathbf{P}_{n+1},\bar{\boldsymbol{\phi}}_{n}^{l},\boldsymbol{\psi}_{n})-\delta\bigl\Vert\boldsymbol{\phi}_{n+1}^{l}-\boldsymbol{\phi}_{n}^{l}\bigr\Vert^{2}$}{\label{line:phiLS:start}

$\boldsymbol{\phi}_{n+1}^{l}=\mathcal{P}_{\phi}(\boldsymbol{\phi}_{n}^{l}-\mu_{n}^{l}\nabla_{\boldsymbol{\phi}^{l}}f(\mathbf{P}_{n+1},\bar{\boldsymbol{\phi}}_{n}^{l},\boldsymbol{\psi}_{n}))$\;

\If{$f(\mathbf{P}_{n+1},\bar{\boldsymbol{\phi}}_{n+1}^{l},\boldsymbol{\psi}_{n})>f(\mathbf{P}_{n+1},\bar{\boldsymbol{\phi}}_{n}^{l},\boldsymbol{\psi}_{n})-\delta\bigl\Vert\boldsymbol{\phi}_{n+1}^{l}-\boldsymbol{\phi}_{n}^{l}\bigr\Vert^{2}$
}{$\mu_{n}^{l}\leftarrow\rho\mu_{n}^{l}$\;}

}}

\For{$k=1$ $\boldsymbol{\mathrm{to}}$ $K$ }{

\Repeat(\tcc*[f]{line search for $\boldsymbol{\psi}^{k}$}){\label{line:psiLS:end}$f(\mathbf{P}_{n+1},\boldsymbol{\phi}_{n+1},\bar{\boldsymbol{\psi}}_{n+1}^{k})\le f(\mathbf{P}_{n+1},\boldsymbol{\phi}_{n+1},\bar{\boldsymbol{\psi}}_{n}^{k})-\delta\bigl\Vert\boldsymbol{\psi}_{n+1}^{k}-\boldsymbol{\psi}_{n}^{k}\bigr\Vert^{2}$}{\label{line:psiLS:start}

$\boldsymbol{\psi}_{n+1}^{k}=\mathcal{P}_{\boldsymbol{\psi}}(\boldsymbol{\psi}_{n}^{k}-\tau_{n}^{k}\nabla_{\boldsymbol{\psi}^{k}}(\mathbf{P}_{n+1},\boldsymbol{\phi}_{n+1},\bar{\boldsymbol{\psi}}_{n}^{k})$\;

\If{$f(\mathbf{P}_{n+1},\boldsymbol{\phi}_{n+1},\bar{\boldsymbol{\psi}}_{n+1}^{k})>f(\mathbf{P}_{n+1},\boldsymbol{\phi}_{n+1},\bar{\boldsymbol{\psi}}_{n}^{k})-\delta\bigl\Vert\boldsymbol{\psi}_{n+1}^{k}-\boldsymbol{\psi}_{n}^{k}\bigr\Vert^{2}$
}{$\tau_{n}^{k}\leftarrow\rho\tau_{n}^{k}$\;}

}}

$n\leftarrow n+1$\;

}

\caption{APGM algorithm for solving (\ref{eq:Opt_problem}). \label{alg:APG:adapmomen}}
}
\end{algorithm}
We remark that (\ref{eq:Opt_problem}) can potentially become a large-scale
optimization problem, and thus, first order methods are particularly
suitable. Indeed, we adopt the \ac{APGM} to derive the proposed method
for solving (\ref{eq:Opt_problem}), which is summarized in \algref{APG:adapmomen}.
Let $(\mathbf{P}_{n},\boldsymbol{\phi}_{n},\boldsymbol{\psi}_{n})$
be the value of $(\mathbf{P},\boldsymbol{\phi},\boldsymbol{\psi})$
at iteration $n$. Then $\mathbf{P}_{n+1}$ is obtained as
\begin{equation}
\mathbf{P}_{n+1}=\mathcal{P}_{P}(\mathbf{P}_{n}-\nu_{n}\nabla_{\mathbf{P}}f(\mathbf{P}_{n},\boldsymbol{\phi}_{n},\boldsymbol{\psi}_{n}))
\end{equation}
where $\nabla_{\mathbf{P}}f(\mathbf{P},\boldsymbol{\phi},\boldsymbol{\psi})$
is the gradient of $f(\mathbf{P},\boldsymbol{\phi},\boldsymbol{\psi})$
\textcolor{black}{\ac{wrt}} $\mathbf{P}^{*}$, $\mathcal{P}_{P}(\cdot)$
denotes the projection onto the set defined by (\ref{eq:P_constr})
and $\nu_{n}$ is the step size, which is found by the line search
routine described in steps \lineref{PLS:start} to \lineref{PLS:end}
of \algref{APG:adapmomen}.

In optimizing the phase shifts of the SIMs, existing algorithms typically
perform a projected gradient step across all phase shifts of all layers
simultaneously using the same step size \cite{an2024two,nadeem2023hybrid}.
However, we have observed that such a method results in poor performance
(see Fig. \ref{fig:CR-and-MI}). To address this, our proposed algorithm
optimizes the phase shifts for the transmit and the receive SIMs on
a layer-by-layer basis, with each layer being assigned a separate
step size. This approach is shown to yield better performance results
in our numerical experiments.

More specifically, the phase shifts of the $l$-th layer of the transmit
SIM at iteration $n+1$ is determined by
\begin{equation}
\boldsymbol{\phi}_{n+1}^{l}=\mathcal{P}_{\boldsymbol{\phi}^{l}}(\boldsymbol{\phi}_{n}^{l}-\mu_{n}^{l}\nabla_{\boldsymbol{\phi}^{l}}f(\mathbf{P}_{n+1},\bar{\boldsymbol{\phi}}_{n}^{l},\boldsymbol{\psi}_{n}))
\end{equation}
for $l=1,2,\ldots,L$, and where we denote $\bar{\boldsymbol{\phi}}_{n}^{l}=[(\boldsymbol{\phi}_{n+1}^{1})^{T},\ldots,(\boldsymbol{\phi}_{n+1}^{l-1})^{T},(\boldsymbol{\phi}_{n}^{l})^{T},\ldots,(\boldsymbol{\phi}_{n}^{L})^{T}]^{T}$
and $\bar{\boldsymbol{\phi}}_{n+1}^{l}=[(\boldsymbol{\phi}_{n+1}^{1})^{T},\ldots,(\boldsymbol{\phi}_{n+1}^{l})^{T},(\boldsymbol{\phi}_{n}^{l+1})^{T},\ldots,(\boldsymbol{\phi}_{n}^{L})^{T}]^{T}$.
In the above, $\nabla_{\boldsymbol{\phi}^{l}}f(\mathbf{P},\boldsymbol{\phi},\boldsymbol{\psi})$
is the gradient of $f(\mathbf{P},\boldsymbol{\phi},\boldsymbol{\psi})$
w.r.t. $\boldsymbol{\phi}^{l*}$, $\mathcal{P}_{\phi^{l}}(\cdot)$
denotes the projection onto (\ref{eq:phi-modulus}), and $\mu_{n}^{l}$
is the step size for the $l$-th layer. The line search procedure
for finding $\mu_{n}^{l}$ is outlined in steps \lineref{phiLS:start}
to \lineref{phiLS:end}. Similarly, $\boldsymbol{\psi}_{n+1}^{k}$
is found as 
\begin{align}
\boldsymbol{\psi}_{n+1}^{k} & =\mathcal{P}_{\boldsymbol{\psi}^{k}}(\boldsymbol{\psi}_{n}^{k}-\tau_{n}^{k}\nabla_{\boldsymbol{\psi}^{k}}f(\mathbf{P}_{n+1},\boldsymbol{\phi}_{n+1},\bar{\boldsymbol{\psi}}_{n}^{k})),
\end{align}
where $\nabla_{\boldsymbol{\psi}^{k}}f(\mathbf{P},\boldsymbol{\phi},\boldsymbol{\psi})$
is the gradient of $f(\mathbf{P},\boldsymbol{\phi},\boldsymbol{\psi})$
w.r.t. $\boldsymbol{\psi}^{k*},$ and $\tau_{n}^{k}$ is the step
size, which is determined in steps \lineref{psiLS:start} to \lineref{psiLS:end}.
The involved gradients are provided in \thmref{gradients}.
\begin{thm}
\label{thm:gradients}The gradients of $f(\mathbf{P},\boldsymbol{\phi},\boldsymbol{\psi})$
with respect to $\mathbf{P}^{*}$, $\boldsymbol{\phi}^{l*}$ and $\boldsymbol{\psi}^{k*}$
are given by
\begin{align}
\nabla_{\mathbf{P}}f(\mathbf{P},\boldsymbol{\phi},\boldsymbol{\psi}) & =-\frac{1}{4\sigma^{2}}\mathbf{H}^{H}\mathbf{H}\ensuremath{\mathbf{P}}\sum\nolimits _{i,j=1}^{N_{vec}}e^{-\frac{F_{i,j}}{4\sigma^{2}}}\APLup\mathbf{x}_{i,j}\APLup\mathbf{x}_{i,j}^{H}\label{eq:grad_P}\\
\nabla_{\boldsymbol{\phi}^{l}}f(\mathbf{P},\boldsymbol{\phi},\boldsymbol{\psi}) & =-\frac{1}{4\sigma^{2}}\sum\nolimits _{i,j=1}^{N_{vec}}e^{-\frac{F_{i,j}}{4\sigma^{2}}}\vect_{d}(\mathbf{L}_{l})\label{eq:grad_phi}\\
\nabla_{\boldsymbol{\psi}^{k}}f(\mathbf{P},\boldsymbol{\phi},\boldsymbol{\psi}) & =-\frac{1}{4\sigma^{2}}\sum\nolimits _{i,j=1}^{N_{vec}}e^{-\frac{F_{i,j}}{4\sigma^{2}}}\vect_{d}(\mathbf{K}_{k})\label{eq:grad_psi}
\end{align}
where 
\begin{subequations}
\begin{align}
\mathbf{L}_{l} & =\boldsymbol{\Theta}^{l+1:L}\mathbf{G}^{H}\mathbf{Z}^{H}\mathbf{H}\ensuremath{\mathbf{P}}\APLup\mathbf{x}_{i,j}\APLup\mathbf{x}_{i,j}^{H}\ensuremath{\mathbf{P}}^{H}\boldsymbol{\Theta}^{1:l-1}(\mathbf{W}^{l})^{H}\\
\mathbf{K}_{k} & =(\mathbf{U}^{k})^{H}\boldsymbol{\Upsilon}^{k-1:1}\mathbf{H}\ensuremath{\mathbf{P}}\APLup\mathbf{x}_{i,j}\APLup\mathbf{x}_{i,j}^{H}\ensuremath{\mathbf{P}}^{H}\ensuremath{\mathbf{B}}^{H}\ensuremath{\mathbf{G}}^{H}\boldsymbol{\Upsilon}^{K:k+1}
\end{align}
\end{subequations}
 where $\boldsymbol{\Theta}^{m:n}=(\mathbf{W}^{m})^{H}(\boldsymbol{\mathbf{\Phi}}^{m})^{H}\cdots(\mathbf{W}^{n})^{H}(\boldsymbol{\mathbf{\Phi}}^{n})^{H}$
and $\boldsymbol{\Upsilon}^{m:n}=(\mathbf{\Psi}^{m})^{H}(\mathbf{U}^{m})^{H}\cdots(\mathbf{\Psi}^{m})^{H}(\mathbf{U}^{n})^{H}$.
\end{thm}
\begin{IEEEproof}
See the Appendix.
\end{IEEEproof}
After calculating the gradients, the next step is to perform a projection
onto the corresponding feasible sets. It is easy to see from (\ref{eq:P_constr})
that $\mathcal{P}_{P}(\mathbf{P})$ is given by
\begin{equation}
\overline{\mathbf{P}}=\mathbf{P}\sqrt{N_{s}/\tr(\mathbf{P}\mathbf{P}^{H})}.
\end{equation}
Since the elements of $\boldsymbol{\phi}^{l}$ are constrained to
lie on the unit circle, $\mathcal{P}_{\phi}(\boldsymbol{\phi}^{l})$
is defined by
\begin{equation}
\overline{\phi_{n}^{l}}=\begin{cases}
\phi_{n}^{l}/\bigl|\phi_{n}^{l}\bigr|, & \phi_{n}^{l}\neq0\\
e^{j\alpha},\alpha\in[0,2\pi] & \phi_{n}^{l}=0.
\end{cases}\label{eq:Proj_phi}
\end{equation}
For $\phi_{n}^{l}=0$, the projection can take any point on the unit
circle. Finally, $\mathcal{P}_{\boldsymbol{\psi}}(\boldsymbol{\psi}^{k})$
is obtained similarly as in (\ref{eq:Proj_phi}).\textcolor{blue}{{}
}\textcolor{black}{The convergence proof of the proposed method follows
the same arguments used in \cite{perovic2022maximum} and is omitted
for brevity. }\vspace{-0.2cm}

\section{Computational Complexity}

In this section, the computational complexity of \algref{APG:adapmomen}
is analyzed by counting the required number of complex multiplications.
We assume that $N\gg N_{t}$ and $E\gg N_{r}$, which is the usual
case for a SIM-based HMIMO communication system. We also assume that
all matrices $\APLup\mathbf{x}_{i,j}\APLup\mathbf{x}_{i,j}^{H}$ are
precomputed.

The complexity of the gradient $\nabla_{\mathbf{P}}f(\mathbf{P},\boldsymbol{\phi},\boldsymbol{\psi})$
is $\mathcal{O}(N_{vec}^{2}N_{s}^{2})$. In addition, $\mathcal{O}(N_{vec}^{2}N_{r}N_{s})$
multiplications are required to obtain $f(\mathbf{P}_{n+1},\boldsymbol{\phi}_{n},\boldsymbol{\psi}_{n})$.
Hence, the complexity of computing the precoding matrix $\mathbf{P}$
is equal~to $\mathcal{O}(I_{p}N_{vec}^{2}N_{r}N_{s})$, where $I_{p}$
is the number of line search~steps.

The complexity of $\sum_{i,j}e^{-\frac{F_{i,j}}{4\sigma^{2}}}\APLup\mathbf{x}_{i,j}\APLup\mathbf{x}_{i,j}^{H}$
in (\ref{eq:grad_phi}) is $\mathcal{O}(N_{vec}^{2}N_{s}^{2})$. Furthermore,
$\mathcal{O}(LN^{3})$ multiplications are needed to calculate $\mathbf{L}_{l}$.
Hence, the complexity of calculating the gradient $\nabla_{\boldsymbol{\phi}^{l}}f(\mathbf{P},\boldsymbol{\phi},\boldsymbol{\psi})$
is $\mathcal{O}(N_{vec}^{2}N_{s}^{2}+LN^{3})$. After obtaining $\boldsymbol{\phi}_{n+1}^{l}$,
the calculation of $\mathbf{B}$ and $\mathbf{H}$ has the complexity
of $\mathcal{O}(\text{\ensuremath{N^{3}}}+ENN_{s})$. The complexity
of calculating $f(\mathbf{P}_{n+1},\bar{\boldsymbol{\phi}}_{n+1}^{l},\boldsymbol{\psi}_{n})$
is $\mathcal{O}(N_{vec}^{2}N_{r}N_{s})$. Therefore, the complexity
of computing the transmit SIM phase shifts $\{\boldsymbol{\phi}^{l}\}_{l=1}^{L}$
can be approximated as $\mathcal{O}(L[LN^{3}+I_{\boldsymbol{\phi}}(\text{\ensuremath{N^{3}}}+ENN_{s}+N_{vec}^{2}N_{r}N_{s})])$,
where $I_{\boldsymbol{\phi}}$ is the number of line search loops.
Similarly, the complexity of computing the receive SIM phase shifts
$\{\boldsymbol{\psi}^{k}\}_{k=1}^{K}$ is equal to $\mathcal{O}(K[KE^{3}+I_{\boldsymbol{\psi}}(\text{\ensuremath{E^{3}}}+ENN_{s}+N_{vec}^{2}N_{r}N_{s})])$,
where $I_{\boldsymbol{\psi}}$ is the number of line search steps.

In summary, the complexity of one iteration of the APGM algorithm
is given by
\begin{gather*}
C_{\text{APGM}}=\mathcal{O}(I_{p}N_{vec}^{2}N_{r}N_{s}+L[LN^{3}+I_{\boldsymbol{\phi}}(\text{\ensuremath{N^{3}}}+ENN_{s}\\
+N_{vec}^{2}N_{r}N_{s})]+K[KE^{3}+I_{\boldsymbol{\psi}}(\text{\ensuremath{E^{3}}}+ENN_{s}+N_{vec}^{2}N_{r}N_{s})]).
\end{gather*}

\section{Simulation Results}

In this section, we evaluate the CR and the MI of \algref{APG:adapmomen}
in a SIM-based HMIMO setup. The channel matrix between the transmit
and the receive SIM is modeled based on the spatially-correlated channel
model as $\mathbf{G}=\mathbf{R}_{\text{R}}^{1/2}\bar{\mathbf{G}}\mathbf{R}_{\text{T}}^{1/2}\in\mathbb{C}^{E\times L}$
\cite{an2023stackedholo,papazafeiropoulos2024achievable} where $\bar{\mathbf{G}}\in\mathbb{C}^{E\times L}$
is distributed according to $\mathcal{CN}(0,\beta\mathbf{I})$; $\beta$
is the free space path loss between the transmit and the receive SIM
modeled as $\beta(d)=\beta(d_{0})+10b\log_{10}(d/d_{0})$, where $\beta(d_{0})=20\log_{10}(4\pi d_{0}/\lambda)$
is the free space path loss at the reference distance $d_{0}$, $b$
is the path loss exponent, and $d$ is the distance between the transmitter
and the receiver. Moreover, $\mathbf{R}_{\text{T}}\in\mathbb{C}^{L\times L}$
and $\mathbf{R}_{\text{R}}\in\mathbb{C}^{E\times E}$ are the spatial
correlation matrices of the transmit and the~receive SIM, respectively,
and their elements are given by \cite{an2023stackedholo}.

In the following simulation setup, the parameters are $\lambda=5\,\text{cm}$
(i.e., $f=6\,\text{GHz}$), $N_{t}=2$, $N_{r}=2$, $N_{s}=2$, $\beta=3.5$,
$d_{0}=1\,\text{m}$, $d=300\,\text{m}$, $L=K=4$ and $\sigma^{2}=-110\,\text{dB}$.
Both the transmit and the receive antennas are placed in arrays parallel
to the $x$-axis, and the midpoints of these arrays have coordinates
$(0,0,0)$ and $(0,0,d)$, respectively. Also, the inter-antenna separations
of these arrays are $\lambda/2$. The meta-atoms in every SIM layer
are uniformly placed in a square formation and the size of each meta-atom
is $\frac{\lambda}{2}\times\frac{\lambda}{2}$. Moreover, all SIM
layers are parallel to the $xy$-plane and their centers are along
the $z$-axis. The separation between the neighboring SIM layers,
as well as the separation between the first SIM layer and the adjacent
antenna array are $\lambda/2$. In the line search procedures in \algref{APG:adapmomen},
all step sizes are initially set to 1000, $\delta=10^{-3}$ and $\rho=1/2$.
The initial values of the optimization variables $\mathbf{P}$, $\boldsymbol{\phi}$,
and $\boldsymbol{\psi}$ are randomly generated. All results are averaged
over 30 independent channel realizations.

\begin{figure}[t]
\centering{}\includegraphics[bb=60bp 568bp 340bp 741bp,clip,width=8.85cm]{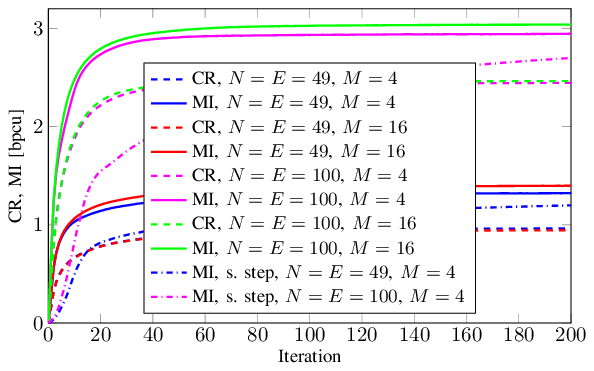}\caption{CR and MI of the proposed APGM algorithm.\label{fig:CR-and-MI}}
\end{figure}

In Fig. \ref{fig:CR-and-MI}, we present the CR and the MI of the
proposed APGM algorithm for different sizes of the discrete symbol
alphabet and different numbers of meta-atoms in SIM layers. In general,
the MI reflects the same trend as observed in the CR, albeit always
achieving larger values. This justifies the use of the CR as an alternative
metric for the MI optimization. In all cases, the CR and the MI reach
90\,\% of their convergent values in approximately 20 iterations
of the proposed algorithm. Also, we notice that the change of $M$
has a negligible influence on the CR, and that the MI may even decrease
slightly if $M$ increases when the transmit and the receive SIMs
have 49 meta-atoms per layer. This phenomenon is due to the reduced
separation between adjacent constellation points when $M$ is increased
without a corresponding increase in the average symbol energy, which
consequently increases the \ac{BEP} per transmission interval. On
the other hand, the CR and the MI demonstrate significant improvements
as the number of meta-atoms in the SIM layers increases. This enhancement
reaffirms the fact that the beamforming capabilities of SIMs are highly
dependent on the number of meta-atoms in SIM layers. \textcolor{black}{To
show the efficacy of the proposed method, we also present the MI obtained
using the same step size for all layers in the SIM. As can be seen
clearly, the corresponding MI is not able to converge even after 200
iterations and is significantly lower than the MI obtained using different
step sizes for each SIM layer as done in our method.}

\begin{figure}[t]
\centering{}\includegraphics[width=8.85cm]{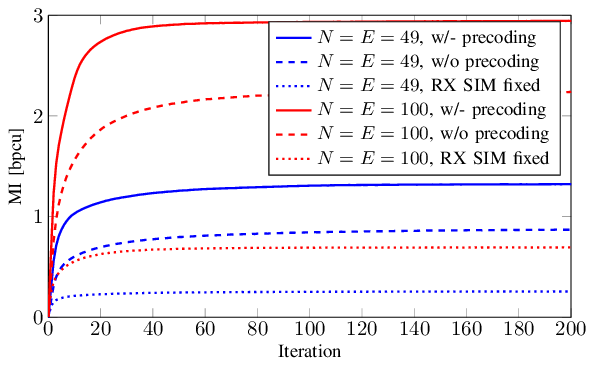}\caption{MI of the considered system with (w/-) and without (w/o) signal precoding
($M=4$). \label{fig:MI_precoder}}
\end{figure}

In Fig. \ref{fig:MI_precoder}, we show the MI of the considered system
with and without signal precoding. It can be clearly observed that
signal precoding can substantially increase the MI for about 47\,\%
and 32\,\% for 49 and 100 meta-atoms per SIM layer, respectively.
On the other side, signal precoding alone, i.e., in the absence of
the transmit and the receive SIMs, is only able to provide a near
zero MI, which is not shown in Fig.~\ref{fig:MI_precoder}.~Hence,
we can conclude that while digital signal precoding of small scale
alone shows limited beamforming gain, but its integration with SIM-based
HMIMO systems can generate a significant impact on the achievable
MI. \textcolor{black}{We also note that the MI is significantly lower
when one of the SIMs employs fixed phase shifts, compared to scenarios
where both SIMs are tunable. This reduction in the MI is approximately
the same regardless of whether the fixed SIM is at the transmit or
the receive end.}\vspace{-0.3cm}

\section{Conclusion}

In this paper, we have demonstrated that the MI in a SIM-based HMIMO
system can be efficiently optimized with the channel CR. To maximize
the CR, we proposed the APGM which optimizes the CR by adjusting the
transmit precoding, and the phase shifts for the transmit and the
receive SIMs on a layer-by-layer basis. Simulation results showed
that the CR is indeed a reliable metric for optimizing the MI in SIM-based
HMIMO systems. Also, integrating even a small scale digital precoder
in the considered system could substantially increase the MI performance.\textcolor{blue}{{}
}\textcolor{black}{For future work, an interesting topic is to study
the potential of SIMs in the context of emerging integrated sensing
and communication systems. Moreover, as a globally optimal solution
has not been reported, it remains to be seen how the performance of
the proposed method is compared to the optimal performance.} \vspace{-0.2cm}

\appendix[Proof of \thmref{gradients}]{}

The gradient of $f(\mathbf{P},\boldsymbol{\phi},\boldsymbol{\psi})$
\textcolor{black}{\ac{wrt}} $\mathbf{P}^{*}$ is given by
\begin{equation}
\!\!\!\nabla_{\mathbf{P}}f(\mathbf{P},\boldsymbol{\phi},\boldsymbol{\psi})=-\frac{1}{4\sigma^{2}}\sum\nolimits _{i,j=1}^{N_{s}}\exp\left(-\frac{F_{i,j}}{4\sigma^{2}}\right)\nabla_{\mathbf{P}}F_{i,j}.\label{eq:grad_P_obj}
\end{equation}
Differentiating $F_{i,j}$, we obtain 
\[
\text{d}F_{i,j}=\tr\{(\mathbf{H}^{H}\mathbf{H}\ensuremath{\mathbf{P}}\APLup\mathbf{x}_{i,j}\APLup\mathbf{x}_{i,j}^{H})^{T}\text{d}\mathbf{P}^{*}+\APLup\mathbf{x}_{i,j}\APLup\mathbf{x}_{i,j}^{H}\ensuremath{\mathbf{P}}^{H}\mathbf{H}^{H}\mathbf{H}\text{d}\mathbf{P}\}
\]
which becomes clear that 
\begin{equation}
\nabla_{\mathbf{P}}F_{i,j}=\mathbf{H}^{H}\mathbf{H}\ensuremath{\mathbf{P}}\APLup\mathbf{x}_{i,j}\APLup\mathbf{x}_{i,j}^{H}.\label{eq:grad_P_F}
\end{equation}
Substituting (\ref{eq:grad_P_F}) into (\ref{eq:grad_P_obj}), we
obtain (\ref{eq:grad_P}).

The gradient of $f(\mathbf{P},\boldsymbol{\phi},\boldsymbol{\psi})$
\textcolor{black}{\ac{wrt}} $\boldsymbol{\phi}^{l*}$ is given by
\begin{equation}
\!\!\!\!\!\!\nabla_{\boldsymbol{\phi}^{l}}f(\mathbf{P},\boldsymbol{\phi},\boldsymbol{\psi})=-\frac{1}{4\sigma^{2}}\sum\nolimits _{i,j=1}^{N_{s}}\exp\left(-\frac{F_{i,j}}{4\sigma^{2}}\right)\nabla_{\boldsymbol{\phi}^{l}}F_{i,j}.\label{eq:grad_Phi_obj}
\end{equation}
Differentiating again $F_{i,j}$ (but now w.r.t. $\boldsymbol{\phi}^{l}$)
yields 
\begin{align}
\text{d}F_{i,j} & =\tr\{\mathbf{L}_{l}\text{d}(\boldsymbol{\mathbf{\Phi}}^{l})^{H}+\mathbf{L}_{l}^{H}\text{d}\boldsymbol{\mathbf{\Phi}}^{l}\}=\tr\{\mathbf{L}_{l}^{T}\text{d}\boldsymbol{\mathbf{\Phi}}^{l*}+\mathbf{L}_{l}^{H}\text{d}\boldsymbol{\mathbf{\Phi}}^{l}\}\nonumber \\
 & =\vect^{T}(\mathbf{L}_{l})\vect(\boldsymbol{\mathbf{\Phi}}^{l*})+\vect^{T}(\mathbf{L}_{l}^{*})\vect(\boldsymbol{\mathbf{\Phi}}^{l}).\label{eq:dF_wrt_Phi}
\end{align}
Thus, we can conclude that
\begin{equation}
\nabla_{\boldsymbol{\phi}^{l}}F_{i,j}=\vect_{d}(\mathbf{L}_{l}),
\end{equation}
and substituting this gradient into (\ref{eq:grad_Phi_obj}), we obtain
(\ref{eq:grad_phi}). Following the same steps, we can also prove
(\ref{eq:grad_psi}).\vspace{-0.3cm}

\bibliographystyle{IEEEtran}
\bibliography{IEEEabrv,IEEEexample,references}

% Generated by IEEEtran.bst, version: 1.14 (2015/08/26)
\begin{thebibliography}{10}
\providecommand{\url}[1]{#1}
\csname url@samestyle\endcsname
\providecommand{\newblock}{\relax}
\providecommand{\bibinfo}[2]{#2}
\providecommand{\BIBentrySTDinterwordspacing}{\spaceskip=0pt\relax}
\providecommand{\BIBentryALTinterwordstretchfactor}{4}
\providecommand{\BIBentryALTinterwordspacing}{\spaceskip=\fontdimen2\font plus
\BIBentryALTinterwordstretchfactor\fontdimen3\font minus
  \fontdimen4\font\relax}
\providecommand{\BIBforeignlanguage}[2]{{%
\expandafter\ifx\csname l@#1\endcsname\relax
\typeout{** WARNING: IEEEtran.bst: No hyphenation pattern has been}%
\typeout{** loaded for the language `#1'. Using the pattern for}%
\typeout{** the default language instead.}%
\else
\language=\csname l@#1\endcsname
\fi
#2}}
\providecommand{\BIBdecl}{\relax}
\BIBdecl

\bibitem{di2020smart}
M.~Di~Renzo \emph{et~al.}, ``Smart radio environments empowered by
  reconfigurable intelligent surfaces: How it works, state of research, and the
  road ahead,'' \emph{IEEE Journal on Selected Areas in Communications},
  vol.~38, no.~11, pp. 2450--2525, 2020.

\bibitem{wang2023energy}
X.~Wang \emph{et~al.}, ``Energy-efficient beamforming for {RISs-aided}
  communications: Gradient based meta learning,'' \emph{arXiv preprint
  arXiv:2311.06861}, 2023.

\bibitem{zhu2023robust}
F.~Zhu \emph{et~al.}, ``Robust {mmWave} beamforming by self-supervised hybrid
  deep learning,'' \emph{arXiv preprint arXiv:2303.12653}, 2023.

\bibitem{lin2022refracting}
Z.~Lin \emph{et~al.}, ``Refracting {RIS-aided} hybrid satellite-terrestrial
  relay networks: Joint beamforming design and optimization,'' \emph{IEEE
  Transactions on Aerospace and Electronic Systems}, vol.~58, no.~4, pp.
  3717--3724, 2022.

\bibitem{zhu2024beamforming}
F.~Zhu \emph{et~al.}, ``Beamforming inferring by conditional {WGAN-GP} for
  holographic antenna arrays,'' \emph{IEEE Wireless Communications Letters},
  vol.~13, no.~7, pp. 2023--2027, 2024.

\bibitem{di2024electromagnetic}
M.~Di~Renzo and M.~D. Migliore, ``Electromagnetic signal and information
  theory,'' \emph{IEEE BITS the Information Theory Magazine}, pp. 1--13, 2024,
  {Early Access}.

\bibitem{an2024exploiting}
K.~An \emph{et~al.}, ``Exploiting multi-layer refracting {RIS}-assisted
  receiver for {HAP-SWIPT} networks,'' \emph{IEEE Transactions on Wireless
  Communications}, 2024, {Early Access}.

\bibitem{an2023stacked}
J.~An \emph{et~al.}, ``{Stacked intelligent metasurface-aided MIMO transceiver
  design},'' \emph{IEEE Wireless Communications}, vol.~31, no.~4, pp. 123--131,
  2024.

\bibitem{liu2022programmable}
C.~Liu \emph{et~al.}, ``A programmable diffractive deep neural network based on
  a digital-coding metasurface array,'' \emph{Nature Electronics}, vol.~5,
  no.~2, pp. 113--122, 2022.

\bibitem{an2024two}
J.~An \emph{et~al.}, ``Two-dimensional direction-of-arrival estimation using
  stacked intelligent metasurfaces,'' \emph{IEEE Journal on Selected Areas in
  Communications}, 2024, {Early Access}.

\bibitem{nadeem2023hybrid}
Q.-U.-A. Nadeem \emph{et~al.}, ``{Hybrid digital-wave domain channel estimator
  for stacked intelligent metasurface enabled multi-user MISO systems},'' in
  \emph{Proc. IEEE Wireless Communications and Networking Conference
  (WCNC)}.\hskip 1em plus 0.5em minus 0.4em\relax IEEE, 2024.

\bibitem{an2023stackedmulti}
J.~An \emph{et~al.}, ``Stacked intelligent metasurfaces for multiuser
  beamforming in the wave domain,'' in \emph{Proc. IEEE International
  Conference on Communications (ICC)}.\hskip 1em plus 0.5em minus 0.4em\relax
  IEEE, 2023, pp. 2834--2839.

\bibitem{an2023stackedholo}
------, ``{Stacked intelligent metasurfaces for efficient holographic MIMO
  communications in 6G},'' \emph{IEEE Journal on Selected Areas in
  Communications}, vol.~41, no.~8, 2023.

\bibitem{papazafeiropoulos2024achievable}
A.~Papazafeiropoulos \emph{et~al.}, ``{Achievable rate optimization for stacked
  intelligent metasurface-assisted holographic MIMO communications},''
  \emph{IEEE Transactions on Wireless Communications}, 2024, {Early Access}.

\bibitem{perovic2018optimization}
N.~S. Perovi{\'c} \emph{et~al.}, ``Optimization of the cut-off rate of
  generalized spatial modulation with transmit precoding,'' \emph{IEEE
  Transactions on Communications}, vol.~66, no.~10, pp. 4578--4595, 2018.

\bibitem{perovic2021optimization}
------, ``{Optimization of RIS-aided MIMO systems via the cutoff rate},''
  \emph{IEEE Wireless Communications Letters}, vol.~10, no.~8, pp. 1692--1696,
  2021.

\bibitem{john2008digital}
J.~G. Proakis and M.~Salehi, \emph{Digital communications}, 5th~ed.\hskip 1em
  plus 0.5em minus 0.4em\relax McGraw-Hill., 2008.

\bibitem{lin2018all}
X.~Lin \emph{et~al.}, ``All-optical machine learning using diffractive deep
  neural networks,'' \emph{Science}, vol. 361, no. 6406, pp. 1004--1008, 2018.

\bibitem{perovic2022maximum}
N.~S. Perovi{\'c} \emph{et~al.}, ``{On the maximum achievable sum-rate of the
  RIS-aided MIMO broadcast channel},'' \emph{IEEE Transactions on Signal
  Processing}, vol.~70, pp. 6316--6331, 2022.

\end{thebibliography}

\end{document}